\documentclass[twoside]{article}

\input ibvs2.sty
\input epsf.sty

\begin{document}

\IBVShead{5618}{23 March 2005}

\IBVStitle{GSC 4232.2830, an Eclipsing Binary with Elliptical Orbit}

\IBVSauth{Goranskij, V. P.$^1$; Shugarov, S. Y\lowercase{u}.$^1$; Kroll, P.$^2$; Golovin, A.$^3$}

\IBVSinst{Sternberg Astronomical Institute, Moscow University, 119992, Russia}
\IBVSinst{Sternwarte Sonneberg, Sternwartestrasse 32, D-96515, Germany}
\IBVSinst{Crimean Astrophysical Observatory, Ukraine, visiting amateur astronomer}

\SIMBADobj{GSC 4232-2830}
\IBVStyp{  EA  }
\IBVSkey{Photometry}

\IBVSabs{The discovery of an Algol type eclipsing variable star having}
\IBVSabs{highly elliptical orbit with the period of 11.628 day is reported.}

\begintext

GSC 4232.2830 (20\hr01\mm28\fsec407, +61\deg10\arcm17\farcs18,
2000.0, $v$=12\mm.1) was suspected to be variable by V.P.G. in
the routine overview of photographic plates taken with 40-cm
astrograph of SAI Crimean station. Two weakenings by
$\approx$0\fmm7 with time difference of 8\fday1 were observed in
1990 September, which suggested that this star was an Algol type
variable star. However, no other eclipses were found in the SAI plate
collection of this sky region including 95 plates taken in 83
nights in the time range JD 2444665--2449358. The analysis of SAI
observations excluded the period 8\fday1, and other possible
periods with P=8\fday1/N (N = 2,3,4,\dots).

To define orbital elements of the binary, we searched for
observations the Sonneberg Observatory plate collection, NSVS
database (Wozniak et al., 2004), and carried out visual monitoring
with a small telescope equipped with an electronic image tube, an
analogue of a night vision device. Later, when we had found a
preliminary solution, we carried out accurate CCD photometry to
improve the orbital elements. A total of our efforts is reflected
in the Table 1. We used the nearby star, GSC 4232.2395 as a
comparison star for CCD photometry, measured its $UBVR_C$
magnitudes relative to V.M.Lyuty's standard near Cyg X-2 (Basko et al.,
1976), and create the uniform standards for eye estimates. The
photometric data for the comparison star, and for the eclipsing
binary GSC 4232.2830 in maximum light are given in Table 2.

We should note, that the depths of eclipses in the NSVS database
do not exceed 0\fmm2, what contradicts to other observations. We
suppose that NSVS measurements concern to integral light of two
stars, a variable star, and a nearby brighter star, GSC
4232.2395, due to low resolution of this survey, that is 72\arcs. The
data given in Table 2 imply the integral $V$ magnitude 11\fmm22,
what is brighter than the NSVS value, 11\fmm68, by 0\fmm46. With this
correction to NSVS magnitudes, and $V$ = 11\fmm70 for GSC~4232.2395, 
we extracted NSVS light curve of the eclipsing binary.

Using all the available observations, we found an orbital
solution with an elliptical orbit and with the period of 11\fday6. The
center of the secondary mimimum occurs at the orbital phase
0\fper69835 $\pm$0.00002 or 8\fday1 after the primary minimum.
The improved ephemeris derived using accurate CCD observations is
the following:
$$ {\rm HJD\ Min\ I} = 2453278.3185(2) + 11\fday628188(5) \times E.$$

\newpage

\begin{table}[hb!]
\begin{center}
{Table 1. The observations of GSC 4232.2830\\} \vspace{0.2cm}
\begin{tabular}{|l|c|c|c|c|c|c|}
\hline
Source                &No.   & ptm     &No.   &Telescope     & Recording     &Observer\\
(J.D. range)          &obs.  & system  &eclip.&              &               &\\
                      &(nights)&       &      &              &               &\\
\hline
Moscow SAI collection & 95   &pg       & 2 & 40-cm, plates   & eye estimates & V.P.G.\\
(2444665-2449358)     &(83)  &         &   &                 &               &       \\
\hline
Sonneberg collection  &262   &pg       & 4?& plates          & eye estimates & S.Yu.Sh.\\
(2426091-2448771)     &(183) &         &   &                 &               &  \\
 \hline
NSVS Database         &525   & V       & 7 & Canon lense     & Thomson       & Wozniak,\\
(2451274-2451630)     &(142) &         &   &                 &TH7899M CCD    & et al.\\
\hline
Nizhny Arkhyz, SAO,   &443   & r       & 3 & 25-cm           & eye estimates & V.P.G.\\
home observatory      &(71)  &         &   & + IP-10         &               &       \\
(2452704-2453279)     &      &         &   & image tube      &               &       \\
\hline
Crimean Observatory   &118   & R$_J$   & 1 & 38-cm           & Apogee-47 CCD & A.G.\\
(2453278)             &(1)   &         &   &                 &               & \\
\hline
Crimean Observatory   &370   &BVR$_J$  & 1 & 38-cm           & Apogee-47 CCD & S.Yu.Sh.\\
(2453321)             &(1)   &         &   &                 &               & \\
\hline
SAI Crimean station   &389   &V        & 1 & 50-cm           & Meade         & S.Yu.Sh.\\
(2453243-2453244)     &(2)   &         &   & Maksutov        & Pictor-416 CCD& \\
\hline
SAI Moscow            &792   &BVR$_J$  & 1 & 70-cm           & Apogee-7p CCD & S.Yu.Sh.\\
(2453263, 2453278)    &(2)   &         &   &                 &               &\\
\hline
Special Astrophysical &118   &UBVR$_C$ & 1 & 100-cm          & EEV42-40 CCD  & V.P.G.\\
Observatory (2453321) &(1)   &         &   &                 &               &\\
\hline
\end{tabular}
\end{center}
\end{table}

\vspace{1cm}

\begin{table}[!hb]
\begin{center}
{Table 2. $UBVR_C$ magnitudes of the nearby (comparison) star, and
out-of-eclipse magnitudes of the eclipsing binary} \vskip2mm
\begin{tabular}{|l|c|c|c|c|}
\hline
Star               & $U$      & $B$      & $V$      & $R_C$ \\
\hline
GSC 4232.2395      & 11\fmm910 & 11\fmm956 & 11\fmm702 & 11\fmm77 \\
                   &$\pm$0.020 &$\pm$0.021 &$\pm$0.028 &$\pm$0.03\\
GSC 4232.2830 (Max)& 13.198    & 12.960    & 12.239    & 11.96 \\
                   &$\pm$0.020 &$\pm$0.018 &$\pm$0.005 &$\pm$0.02 \\
\hline
\end{tabular}
\end{center}
\end{table}

\vspace{1cm}

\begin{table}[!hb]
\begin{center}
{Table 3. Times of Minima} \vskip2mm
\begin{tabular}{|lcrl|lcrl|}
\hline
JD hel.    & Min&   $O-C$   & Obs. &  JD hel.    & Min&   $O-C$   & Obs. \\
2400000+   &    &   day   & set  &  2400000+   &    &   day   & set  \\
\hline
31204.512: & II &  0.0019 & Sonneberg   & 51452.68   & I  & $-$0.0130 & NSVS \\
31739.392: & II & $-$0.0148 & Sonneberg   & 51487.60   & I  &  0.0225 & NSVS \\
37960.505  & II &  0.0176 & Sonneberg   & 51603.95   & I  &  0.0906 & NSVS \\
38673.319  & I  &  0.0046 & Sonneberg   & 52906.303  & I  &  0.0864 & it   \\
48150.318  & I  &  0.0304 & SAI         & 53150.41   & I  &  0.0016 & it   \\
48158.427  & II &  0.0188 & SAI         & 53193.413  & II & $-$0.0006 & it   \\
51324.758  & I  & $-$0.0249 & NSVS        & 53243.4335 & I  & $-$0.0004 & CCD  \\
51359.73   & I  &  0.0625 & NSVS        & 53263.1827 & II & $-$0.0001 & CCD  \\
51382.86   & I  & $-$0.0639 & NSVS        & 53278.3185 & I  &  0.0000 & CCD  \\
51402.69   & II &  0.0225 & NSVS        & 53321.3237 & II &  0.0000 & CCD  \\
\hline
\end{tabular}
\end{center}
\end{table}

\newpage
\IBVSfig{12cm}{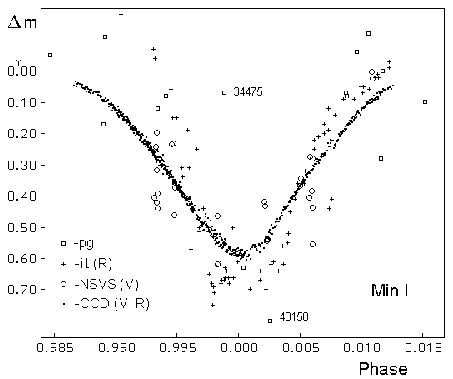}{Observations of GSC 4232.2830 in the primary minima.}
\IBVSfigKey{5618-f1.eps}{GSC 04232-02830}{light curve}

The moments of weakenings and mid-eclipses are given in Table 3.
$O-C$ analysis does not show orbital period variations during the
time interval of observations, or any evidence of the apsidal
motion.

The results of all the observations are shown in Fig.~1 for
Min I, and in Fig.~2 for Min II. The magnitudes in different
filters are calculated relative to out-of-eclipse level, and
combined together with small shifts along the magnitude axis, if
needed.

The observations show that both eclipses have about equal depth,
$\approx$0\fmm60, but essentially different duration, 0\fper028
(7\fhr8) for Min I, and 0\fper0175 (4\fhr9) for Min II. The
eclipses are partial. Using the displacement of the secondary
minimum and eclipse width ratio, we calculate the orbital
eccentricity of 0.39, and $\omega$ = 322\deg. CCD photometry gives
mean colours $U-B$ = 0\fmm238$\pm$0\fmm027, and $B-V$ = 0\fmm721$\pm$0\fmm019 
without notable colour variations in the eclipse
phases. These colours suggest that the components of the system
are solar type main sequence stars.

We used the same set and magnitudes of comparison stars to reduce
the photographic eye estimates. The old Sonneberg
photographic observations indicate that the eclipses were
shallower in the middle of the past century than in the present
time. There are some contradictions between observations marked
in Fig.~1 and 2, when the observer does not notice
weakening in the eclipse phases. One photographically traced
eclipse, and 2-2 outstanding data points are marked with their
Julian dates (JD$-$2400000) in these Figures.

\newpage

\IBVSfig{12cm}{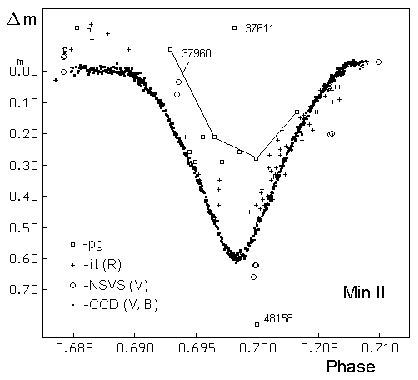}{Observations of GSC 4232.2830 in the secondary minima}
\IBVSfigKey{5618-f2.eps}{GSC 04232-02830}{curve}

The contradictions may suggest that the depth of eclipses varied,
as in the well known system SS Lac (Mossakovskaya, 1993; Milone
et al, 2000; Torres and Stefanik, 2001). The eclipse depth
variations should be verified with more precise observations
taken during a longer time interval.

\references

Basko, M.M., Goranskij, V.P., Lyuty, V.M., et al., 1976,
{\it Variable Stars}, {\bf 20}, 219

Milone, E.F., Schiller, S.J., Munari, U. \& Kallrath,
J, 2000, {\it AJ}, {\bf 119}, 1405

Mossakovskaya, L.V., 1993, {\it Astron. Letters}, {\bf 19}, 35

Torres, G. \& Stefanik, R.P., 2000, {\it AJ}, {\bf 119}, 1914

Wozniak, P.R., Vestrand, W.T., Akerlof, C.W., et al., 2004, {\it AJ}, {\bf 127}, 2436

\endreferences

\end{document}